 \documentclass[nohyper,12pt,letterpaper]{JHEP3}
 \usepackage{epsfig}
 
 \newcommand{\vev}[1]{{\left< {#1} \right>}} 

\title{Exact results for static and radiative fields of a quark in ${\cal N}=4$ super Yang-Mills}

\author{Bartomeu Fiol$^1$, Blai Garolera$^1$ and Aitor Lewkowycz$^2$  \\

$^1$Departament de F{\'\i}sica Fonamental i \\Institut de Ci{\`e}ncies del Cosmos, 

Universitat de Barcelona,

Mart{\'\i}\ i Franqu{\`e}s 1, 08028 Barcelona, Catalonia, Spain \\

$^2$Perimeter Institute for Theoretical Physics,

Waterloo, Ontario N2L 2Y5, Canada \\

\email{bfiol@ub.edu, bgarolera@ffn.ub.es, alewkowycz@pitp.ca}}

\abstract{In this work (which supersedes our previous preprint \cite{Fiol:2011aa}) we determine the expectation value of the ${\cal N}=4$ SU(N) SYM Lagrangian density operator in the presence of an infinitely heavy static particle in the symmetric representation of SU(N), by means of a D3-brane probe computation. The result that we obtain coincides with two previous computations of different observables, up to kinematical factors. We argue that these agreements go beyond the D-brane probe approximation, which leads us to propose an exact formula for the expectation value of various operators. In particular, we provide an expression for the total energy loss by radiation of a heavy particle in the fundamental representation.}  

\begin{document}

\section{Introduction}
Exact results for generic 4d quantum field theories are extremely hard to come by. The situation improves when one considers quantum field theories with additional symmetries, as conformal invariance and/or supersymmetry, since these additional symmetries constrain the parametric dependence of a variety of interesting quantities, that can sometimes be determined exactly. 

One of the most intensively studied theories with such additional symmetries is ${\cal N}=4$ SYM, which is both conformally invariant and maximally supersymmetric. Among its local gauge invariant operators, one encounters the chiral primary operators (CPOs) and their descendants, which fall into short multiplets of the superconformal algebra and enjoy various special properties \cite{Dolan:2002zh, Lee:1998bxa, Baggio:2012rr}. For the purposes of this work, we will be chiefly interested in the supercurrent multiplet, whose CPO has scaling dimension $\Delta=2$, since both the Lagrangian density and the stress-energy tensor belong to this multiplet.

Among the non-local gauge invariant operators of ${\cal N}=4$ SYM, locally BPS Wilson loops and Wilson lines have also been intensively studied over the years. They are characterized by a contour in space-time and  a representation of the gauge group, and their vacuum expectation value has been computed exactly in a few cases \cite{Drukker:2000rr, Drukker:2006ga}. Finally, there have been a number of works devoted to computing correlation functions of Wilson loops with local operators \cite{Semenoff:2001xp, Okuyama:2006jc, Giombi:2006de}.

In a superficially different line of research, the behavior of external probes and the response of the fields to such probes, both in vacuum and at finite temperature, have been intensively scrutinized using the AdS/CFT duality \cite{Aharony:1999ti}. One first goal of the present note is to continue the study of such computations, by presenting the evaluation of the one-point function of the Lagrangian density in the presence of a static heavy probe in the symmetric representation\footnote{This computation appeared in our previous preprint \cite{Fiol:2011aa}, which has been superseded by the present work.}. A second, and in our opinion, farther reaching goal is to argue that some of the questions that appear in the study of external probes can be answered exactly, by relating them to the evaluation of certain correlation functions involving Wilson loops and local operators. To be specific, we propose that from the two-point function of the circular Wilson loop in the fundamental representation and the $\Delta=2$ CPO, computed exactly in \cite{Okuyama:2006jc} (and normalized by the vev of the circular Wilson loop computed in \cite{Drukker:2000rr}) one can read off the exact one-point functions of this CPO and (more importantly) its descendants, in the presence of a heavy particle following either a static trajectory (straight Wilson line) or a trajectory with constant proper acceleration (hyperbolic Wilson line). Up to kinematic factors, these one-point functions are given by the following function  
\begin{equation}
f(\lambda,N)=\frac{\lambda}{64 \pi^2N}\frac{L_{N-1}^2(-\frac{\lambda}{4N})+L_{N-2}^2(-\frac{\lambda}{4N})}{L_{N-1}^1(-\frac{\lambda}{4N})}
\label{ourctwo}
\end{equation}
where $L_n^\alpha$ are generalized Laguerre polynomials.

The structure of this note is as follows. In section 2 we review the study of external probes within the framework of the AdS/CFT correspondence, and the computation of various one-point functions of operators in the presence of such external probes. In section 3 we present\footnote{See footnote 1.} the computation of the one-point function of the Lagrangian density for a static heavy particle in the symmetric representation, by means of the study of the perturbation of the dilaton profile caused by a certain D3-brane in the $AdS_5$ background. We note that the result that we obtain is, up to the respective kinematical factors, exactly the same as in two previous D3-brane probe computations that have appeared in the literature, a first one regarding the two-point function of a circular Wilson loop with a particular chiral primary operator \cite{Giombi:2006de}\footnote{We would like to thank N. Drukker for pointing out this reference to us, and for urging us to compare our results with the ones that appear there.}, and a second one devoted to the computation for energy loss by radiation of a heavy particle in the symmetric representation \cite{fiol11}. Finally, in section 4 we address why the results of these different computations ought to coincide, and argue that their agreement holds for arbitrary representations of the gauge group and beyond the D-brane probe approximation. This leads us to propose that existing exact results \cite{Okuyama:2006jc} provide an exact formula for all those quantities.

{\bf Note Added:} Quite recently, a very interesting preprint appeared \cite{correa} presenting arguments, complementary to those provided here, for an exact formula for the radiation of a moving quark in ${\cal N}=4$ SYM. While the possible connections between their and our arguments remain to be fully sorted out, happily the proposed formulas for energy loss exactly agree. Indeed, their formula depends on a function $B(\lambda,N)$
$$
B(\lambda,N)=\frac{1}{2\pi^2}\lambda \partial_\lambda \log \vev{W}
$$
where $\vev{W}$ is the vev of the circular Wilson loop obtained by a special conformal transformation of the 1/2 BPS straight line, and computed exactly in \cite{Drukker:2000rr}. Considering the explicit form of $\vev{W}$ \cite{Drukker:2000rr}, it is a simple matter to check that this function $B$ is up to a numerical factor our function $f$, eq. (\ref{ourctwo}). In fact, using that 
$$
\partial_\lambda L_{N-1}^1(-\frac{\lambda}{4N})=\frac{1}{4N}L_{N-2}^2(-\frac{\lambda}{4N})
$$
and $L_{N-1}^1=L_{N-1}^2-L_{N-2}^2$ one easily sees that $B=4f$, so the dependence on $\lambda$ and $N$ is exactly the same.

\section{External probes in $AdS/CFT$}
One of the many applications of the AdS/CFT correspondence is the study of the behavior of external probes and the response of the fields to the presence of those probes. The first example of such computations was the evaluation of the static quark-antiquark potential in \cite{Rey:1998ik, Maldacena:1998im} by means of a particular string configuration reaching the boundary of $AdS$. Following those seminal works, the key idea of realizing external heavy quarks by strings in the bulk geometry has been generalized in many directions. In particular, as we will briefly review, probes transforming under different representations of the gauge group are holographically realized by considering different types of branes in the supergravity background.

An external probe in the fundamental representation of the gauge group is dual to a string in the bulk. At least for the simplest implementations of this identification (i.e. in the absence of additional scales like finite mass or non-zero temperature), the computed observables reveal a common feature: if we identify at weak coupling $\lambda$ as the analogous of the charge squared, at strong coupling there is a screening of this charge, in the sense that the results obtained are similar to the ones we would obtain in classical electrodynamics, but with the strong coupling identification
\begin{equation}
e^2_{\Box} \sim \sqrt{\lambda}
\label{chargefund}
\end{equation}
This generic behavior stems from the fact that the Nambu-Goto action evaluated for world-sheet metrics embedded in $AdS_5$ goes like
$$
S_{NG}=-\frac{1}{2\pi \alpha'} \int d^2 \sigma \sqrt{-|g|} =-\frac{\sqrt{\lambda}}{2\pi L^2}\int d^2 \sigma \sqrt{-|g|}
$$
where $L$ is the $AdS_5$ radius which generically cancels out from this expression when specific world-sheet metrics are plugged-in.  Some examples of this are the original quark-antiquark potential \cite{Rey:1998ik, Maldacena:1998im},  the expectation value of gauge invariant operators in the presence of  a particle at rest \cite{hep-th/9812007, hep-th/9906153} or following arbitrary motion \cite{Athanasiou:2010pv, arXiv:1106.4059}, and the formula for energy loss by radiation \cite{hep-th/0305196}.

This leading $\sqrt{\lambda}$ result is expected to receive $1/N$ and $1/\sqrt{\lambda}$ corrections. The computation of $1/\sqrt{\lambda}$ corrections is addressed for instance in \cite{Forste:1999qn}. As it turns out, a possible venue to compute results that capture $1/N$ corrections is to switch to probes transforming in higher rank representations of the gauge group. It is by now well understood that on the gravity side these probes are realized by D3 and D5 branes. Specifically, the duals of particles in the symmetric or antisymmetric representations of the gauge group are given by D3 and D5 branes respectively, with world-volume fluxes that encode the rank of the representation \cite{Gomis:2006sb, Yamaguchi:2006tq, Hartnoll:2006hr}. One of the novel features of this identification is that some computed observables are functions of $k/N$, where $k$ is the rank of the symmetric/antisymmetric representation. This allows to explore the AdS/CFT correspondence beyond the leading large $N$, large $\lambda$ regime. 

While the holographic prescription is in principle equally straightforward for the study of probes in the symmetric and the antisymmetric representations, when it comes to actual computations we currently face more difficulties in the symmetric case than in the antisymmetric one. One of the reasons behind this difference comes from the existence of a quite universal result for the embedding of D5 branes in terms of embeddings of fundamental strings, due to Hartnoll \cite{Hartnoll:2006ib}: given a string world-sheet that solves the Nambu-Goto action on an arbitrary Ricci flat manifold $M$, there is a quite general construction that provides a solution for the D5-brane action in $M\times S^5$, of the form $\Sigma \times S^4$ where $\Sigma \hookrightarrow M$ is the string world-sheet and $S^4\hookrightarrow S^5$. This gives a link between the string used to describe a particle in the fundamental representation and the D5 brane used to represent a probe in the antisymmetric representation. Moreover,
\begin{equation}
\sqrt{\lambda} \rightarrow \frac{2N}{3\pi}\sin^3 \theta_k \sqrt{\lambda} \sim  e^2_{A_k}
\label{chargeanti}
\end{equation}
where $\theta_k$ denotes the angle of $S^4$ inside $S^5$ and is the solution of \cite{hep-th/0104082}
$$
\sin \theta_k \cos \theta_k -\theta_k=\pi\left(\frac{k}{N}-1\right)
$$
This identification is supported by explicit computations of Wilson loops \cite{Yamaguchi:2006tq,Hartnoll:2006ib} which match matrix model computations \cite{Hartnoll:2006is}, energy loss by radiation in vacuum \cite{fiol11} and in a thermal medium \cite{Chernicoff:2006yp}, or the impurity entropy in supersymmetric versions of the Kondo model \cite{Mueck:2010ja}.

On the other hand, for probes in the symmetric representation we currently don't have a generic construction that links the string that realizes a particle in the fundamental representation with a D3 brane that realizes a probe in the symmetric representation. Furthermore, while the observables analyzed so far in the symmetric representation seem to depend on the combination
\begin{equation}
\kappa=\frac{k\sqrt{\lambda}}{4N}
\label{defkappa}
\end{equation}
introduced in \cite{Drukker:2005kx}, they do not display a common function that replaces the $\sqrt{\lambda}$ dependence of the fundamental representation. For instance, in the computation of the energy loss by radiation in vacuum of a particle moving with constant proper acceleration, it was found in \cite{fiol11} that
\begin{equation}
\sqrt{\lambda} \rightarrow 4N\kappa \sqrt{1+\kappa^2} = k\sqrt{\lambda}\sqrt{1+\frac{k^2 \lambda}{16 N^2}} \stackrel{?}{\sim} e^2_{S_k}
\label{chargesym}
\end{equation}
while for the vev of a circular Wilson loop, it was found that \cite{Drukker:2005kx}
$$
\sqrt{\lambda} \rightarrow 2N (\kappa \sqrt{1+\kappa^2}+ \sinh^{-1} \kappa)
$$

While both functions expanded as a power series  in $\kappa$ start with the common term $k\sqrt{\lambda}$ (i.e. $k$ times the result for the fundamental representation) they are clearly different beyond this leading order.

In order to shed some light on the issue of observables for probes in the symmetric representation, in the next section we will compute the expectation value of the Lagrangian density in the presence of an infinitely heavy half-BPS static particle, transforming in the k-symmetric representation of ${\cal N}=4$ $SU(N)$ SYM. This operator is sourced by the asymptotic value of the dilaton \cite{hep-th/9702076} (we follow the conventions of \cite{arXiv:1106.4059}) , 
$$
{\cal O}_{F^2}=\frac{1}{2 g_{YM}^2} \mbox{Tr }\left(F^2 +[X_I,X_J][X^I,X^J] \; \hbox{+ fermions }\right)
$$
On general grounds, in the presence of a static probe placed at the origin, and transforming in the ${\cal R}$ representation of the gauge group, the one-point function will be of the form
$$
\vev{{\cal O}_{F^2}(\vec x)}_{\cal R}=\frac{f_{\cal R}(\lambda, N)}{|\vec x|^4}
$$ 
and our objective is to compute the  dimensionless function $f_{{\cal S}_k}(\lambda, N)$ when the probe transforms in the k-symmetric representation of $SU(N)$. By analogy with the Coulombic case, one might refer to $f_{{\cal S}_k}(\lambda, N)$ as the square of the ``chromo-electric'' charge of the heavy particle. To carry out this computation we will consider a particular half-BPS D3-brane embedded in $AdS_5\times S^5$ and analyze the linearized perturbation equation for the dilaton, with the D3-brane acting as source. The advantage of considering this operator is that the perturbation equation of the dilaton decouples from the equations for the metric and RR field perturbations, so its study is quite straightforward.

The analogous computation for a particle in the fundamental representation was carried out some time ago, considering in that case the perturbation equation for the dilaton sourced by a fundamental string \cite{hep-th/9812007, hep-th/9906153}. For future reference, let's end this section by quoting their final result in our conventions,
\begin{equation}
\vev{{\cal O}_{F^2}(\vec x)}_{\Box}\; = \; \frac{\sqrt{\lambda}}{16\pi^2}\frac{1}{|\vec x|^4}
\label{onepointfund}
\end{equation}

\section{Static fields via a probe computation}
In this section we will present the details of the computation of $\vev{{\cal O}_{F^2}}$ in the presence of a static heavy probe transforming in the symmetric representation. We will first compute the linearized perturbation of the dilaton field caused by the D-brane probe, and from its behavior near the boundary of $AdS_5$ we will then read off the vev of  ${\cal O}_{F^2}$. Our computations will closely follow the ones presented in \cite{hep-th/9812007, hep-th/9906153} for the case of a probe in the fundamental representation.

We work in Poincar\'e coordinates and take advantage of the spherical symmetry of the problem
$$
ds^2_{AdS_5}=\frac{L^2}{z^2}\left(dz^2-dt^2+dr^2+r^2 d\theta^2+r^2\sin^2 \theta d\varphi^2\right)
$$
The D3-brane we will be interested in was discussed in \cite{Rey:1998ik,Drukker:2005kx}. It reaches the boundary of AdS ($z=0$ in our coordinates) at a straight line $r=0$, which is the world-line of the static dual particle placed at the origin. Since we let the D3 brane reach the boundary, the static particle is infinitely heavy. To describe the D3-brane, identify $(t,z,\theta,\varphi)$ as the world-volume coordinates. Then the solution is given by
$$
r=\kappa z\hspace{1cm}F_{tz}=\frac{\sqrt{\lambda}}{2\pi}\frac{1}{z^2}
$$
with $\kappa$ as defined in eq. (\ref{defkappa}). As shown in detail in \cite{Drukker:2005kx} this D3-brane is $1/2$-BPS. 

Our next step is to consider at linear level the backreaction that this D3-brane induces on the $AdS_5\times S^5$ solution of IIB supergravity. More specifically, since the dilaton is constant in the unperturbed solution, and its stress-energy tensor is quadratic, at the linearized level the equation for the perturbation of the dilaton decouples from the rest of linearized supergravity equations. As in \cite{hep-th/9812007, hep-th/9906153}, we work in Einstein frame, and take as starting point the action
$$
S=-\frac{\Omega_5L^5}{2\kappa_{10}^2}\int d^5x \sqrt{-|g_E|}\frac{1}{2}g_E^{mn}\partial_m \phi \partial_n \phi
-T_{D3}\int d^4\xi \sqrt{-|G_E+e^{-\phi/2}2\pi \alpha' F|}
$$
The resulting equation of motion can be written
$$
\partial_m\left(\sqrt{-|g_E|} g_E^{mn}\partial _n \phi\right)=J(x)
$$
with the source defined by the D3-brane solution
$$
J(x)=\frac{T_{D3} \kappa_{10}^2}{\Omega_5 L} \frac{\kappa \sin \theta}{z^2} \delta\left(r-z\kappa\right)
$$
To compute $\phi(x)$ we will use its Green function $D(x,x')$
$$
\phi(x)=\int d^5x'\; D(x,x')\; J(x')
$$
It is convenient to write $D(x,x')$ purely in terms of the invariant distance $v$ defined by
\begin{equation}
\cos v=1-\frac{(t-t')^2-(\vec x-\vec x')^2-(z-z')^2}{2zz'}
\label{invav}
\end{equation}
The explicit expression for $D(v)$ can be found for instance in \cite{hep-th/9812007}
$$
D=\frac{-1}{4\pi^2L^3 \sin v}\frac{d}{dv}\left(\frac{\cos 2v}{\sin v} \Theta(1-|\cos v|)\right)
$$
To carry out the integration, we follow the same steps as \cite{hep-th/9812007}. We first define a rescaled dilaton field,
$$
\tilde \phi\equiv \frac{\Omega_5 L^8}{2\kappa_{10}^2}\phi
$$
and use eq. (\ref{invav}) to change variables from $t'$ to $v$ to obtain, after an integration by parts
$$
\tilde \phi=\frac{N\kappa z^2}{16 \pi^4}\int _0^\infty dr'  \int_0^\pi d\theta' \sin \theta'\int_0^{2\pi} d\varphi' \int _0^\infty \frac{dz' \; \delta(r'-z'\kappa)}{\left(z^2+z'^2+(\vec x-\vec x')^2\right)^{\frac{3}{2}}}\int _0 ^\pi \frac{dv\; \cos 2v}{\left(1-\frac{2zz'\; \cos v}{z^2+z'^2+(\vec x-\vec x')^2}\right)^{\frac{3}{2}}} 
$$
The integral over $v$ is the same one that appeared in the computation of the perturbation caused by a string dual to a static probe \cite{hep-th/9906153}. The novel ingredient in the computation comes from the non-trivial angular dependence in the current case. While it might be possible to completely carry out this integral, at this point it is pertinent to recall that to compute the expectation value of the dual field theory operator, we only need the leading behavior of the perturbation of the dilaton field near the boundary of $AdS_5$. Specifically \cite{hep-th/9812007}, 
\begin{equation}
\vev{{\cal O}_{F^2}}=-\frac{1}{z^3}\partial_z \tilde \phi |_{z=0}
\label{onepoint}
\end{equation}
so for our purposes it is enough to expand the integrands in powers of $z$, and keep only the leading $z^4$ term. This simplifies the task enormously, and reduces it to computing some straightforward integrals. Skipping some unilluminating steps we arrive at
$$
\tilde \phi= \frac{N\kappa}{16 \pi^2}\frac{z^4}{(z^2+r^2)^2}\frac{1}{(1+\kappa^2)^{3/2}}\frac{1}{\left(1-\frac{\kappa^2}{1+\kappa^2}\frac{r^2}{r^2+z^2}\right)^2}+O(z^5)
$$
which upon differentiation, and setting then $z=0$ as required by eq. (\ref{onepoint}), leads to our final result
\begin{equation}
\vev{{\cal O}_{F^2}}_{S_k}=\frac{N\kappa \sqrt{1+\kappa^2}}{4\pi^2}\frac{1}{|\vec x|^4}=\frac{k\sqrt{\lambda}}{16\pi^2}\frac{\sqrt{1+\frac{k^2 \lambda}{16 N^2}}}{|\vec x|^4}
\label{ourresult}
\end{equation}
to be compared with the result for the fundamental representation (\ref{onepointfund}).
 
\section{Exact results for static and radiative fields}
The coefficient that appears in the one-point function (\ref{ourresult}) computed in the previous section has appeared before in the literature, at least in two different computations. Let us review them in turns. The first place where this coefficient appears is in the computation of the large distance behavior of the correlation function of a circular Wilson loop in the symmetric representation with the $\Delta=2$ chiral primary operator by means of a D3-brane \cite{Giombi:2006de}. A circular Wilson loop can be expanded in terms of local operators when probed from distances much larger than its radius, and the coefficients appearing in this OPE can be read off from the large distance behavior of the two-point function of the Wilson loop and the local operators\cite{Berenstein:1998ij}
$$
\frac{\vev{W(C){\cal O}_n}}{\vev{W(C)}}=c_n\frac{R^{\Delta_n}}{L^{2\Delta_n}}+\dots
$$
The authors of \cite{Giombi:2006de} computed the coefficients $c_n$ in the case that the local operators are chiral primary operators given by symmetric traceless combinations of scalar fields, and for a Wilson loop in the symmetric or antisymmetric representation. For the symmetric representation and in their normalization\footnote{\cite{Giombi:2006de,Berenstein:1998ij} consider CPOs whose two-point function is unit-normalized \cite{Lee:1998bxa}.} for ${\cal O}_n$, they obtained (eq. (4.20) in \cite{Giombi:2006de})
$$
c_{S_k,\Delta}^{GRT}=\frac{2^{\Delta/2+1}}{\sqrt{\Delta}}\sinh (\Delta \sinh^{-1} \kappa)
$$
In the previous section we computed a one-point function for the Lagrangian density, which belongs to the same short multiplet as the chiral primary operator with $\Delta=2$, so we are interested in comparing our result with the previous formula for $\Delta=2$
\begin{equation}
c_{S_k,2}^{GRT}=4\sqrt{2}\kappa\sqrt{1+\kappa^2}
\label{cgrt}
\end{equation}
which displays the same dependence in $\lambda$ and $N$ as our result (\ref{ourresult}), except for the overall differing normalization conventions for ${\cal O}_2$ (as a check, both results reproduce in the corresponding limit the previously known results for the fundamental representation, $k=1$). 

We are now going to argue that this agreement can be understood as following from generic properties of ${\cal N}=4$ SYM, and it is a reflection of exact relations among expectation values of various operators. These relations are valid for any representation of the gauge group, and beyond the regime of validity of the D-brane probe computation. Our argument proceeds in three steps: first, we notice that the one-point function of an operator in the presence of an external probe is equivalent to the two-point function of the operator with the corresponding Wilson loop, normalized by the vev of the Wilson loop.

As the second step in our argument, we claim that the coefficient that appears in the normalized two-point function of a circular Wilson loop with a CPO is the same that the one that appears in the (trivially, since $\vev{W}=1$ for the straight line) normalized two-point function of a straight line Wilson loop with the same CPO. One argument\footnote{due to N. Drukker.} is that in the conformal transformation from the line to the circle, both the two-point function and the vev of the Wilson loop pick up the same anomalous contribution, so it will cancel in the ratio.

Finally, since the Lagrangian density and the stress-energy tensor are descendants of ${\cal O}_2$, we expect that their normalized two-point functions with the circular Wilson loop are determined by the same coefficient as the normalized two-point function of ${\cal O}_2$.

This line of reasoning explains the agreement found between results (\ref{ourresult}) and (\ref{cgrt}), obtained in the D-brane probe approximation for external sources in the symmetric representation, but goes well beyond this particular case. If we are somehow able to exactly compute the normalized two-point function of ${\cal O}_2$ and the $1/2$ BPS circular Wilson loop in some representation, we claim that this also gives the exact one-point function of the Lagrangian density or the stress-energy tensor in the presence of a heavy particle in that representation. Luckily, some of the relevant computations have already been performed; for instance, in \cite{Okuyama:2006jc}, Okuyama and Semenoff computed the exact two-point function of a circular Wilson loop in the fundamental representation and a chiral primary operator, by means of a normal matrix model. In particular, for the unit-normalized chiral primary operator ${\cal O}_2$ they obtain (eq. (4.5) in \cite{Okuyama:2006jc}) 
$$
\vev{W(C)_\Box{\cal O}_2}=\frac{\sqrt{2} \lambda}{4N^3}\left[L_{N-1}^2(-\frac{\lambda}{4N})+L_{N-2}^2(-\frac{\lambda}{4N})\right]e^{\frac{\lambda}{8N}} 
$$
so after normalizing by the vev of the circular Wilson loop computed in \cite{Drukker:2000rr} we obtain
$$
c_2=\frac{\vev{W(C)_\Box{\cal O}_2}}{\vev{W(C)_\Box}}=\frac{\sqrt{2}\lambda}{4N^2}\frac{L_{N-1}^2(-\frac{\lambda}{4N})+L_{N-2}^2(-\frac{\lambda}{4N})}{L_{N-1}^1(-\frac{\lambda}{4N})}
$$
The authors of \cite{Giombi:2006de} already showed that this matrix model result reduces in the corresponding limit to the one computed with the D3-brane probe approximation, eq. (\ref{cgrt}). Furthermore, in the planar limit it reproduces the result of \cite{Semenoff:2001xp}. Applying the reasoning presented above, we therefore propose that the exact one-point function of ${\cal O}_{F^2}$ in the presence of a heavy probe in the fundamental representation is given by
\begin{equation}
\vev{{\cal O}_{F^2}(\vec x)}_{\Box}=\frac{\lambda}{64\pi^2 N}\frac{L_{N-1}^2(-\frac{\lambda}{4N})+L_{N-2}^2(-\frac{\lambda}{4N})}{L_{N-1}^1(-\frac{\lambda}{4N})}
\frac{1}{|\vec x|^4}
\label{resultone}
\end{equation}
This is one of our main results. It reduces in the large $\lambda$, large $N$ limit to the known result (\ref{onepointfund}). More than that, while the range of validity of the probe computation in the previous section a priori does not include setting $k=1$ (i.e. considering the fundamental representation) if we nevertheless go ahead and set $k=1$, it can be checked that (\ref{ourresult}) correctly captures the large $\lambda$, large $N$ limit with $\kappa$ fixed of (\ref{resultone}). This type of agreement between a matrix model result and a D3-brane computation, beyond the expected regime of validity of the D-brane probe approximation, has been observed before \cite{Drukker:2005kx}.

So far we have been discussing static sources. Let's now turn to particles undergoing accelerated motion, where we will encounter for the second time that the coefficient of the one-point function (\ref{ourresult}) computed in the previous section had appeared before in the literature. In \cite{fiol11}, two of the present authors computed the energy loss by radiation of an infinitely heavy particle transforming in the symmetric or antisymmetric representation, in the large $\lambda, N$, fixed $\kappa$ limit. For the symmetric representation, it was possible to carry out the computation only for the particular case of a particle undergoing motion with constant proper acceleration, and the total radiated power obtained was \cite{fiol11}
\begin{equation}
P_{S_k}=\frac{2N\kappa}{\pi}\sqrt{1+\kappa^2}\frac{1}{R^2}
\label{powersym}
\end{equation}
where $R$ appears in the hyperbolic trajectory $-(x^0)^2+(x^1)^2=R^2$. Note that again, apart from kinematic factors, the coefficient that appears in the total radiated power, eq. (\ref{powersym}), is the same one as in the one-point function computed in the previous section, eq. (\ref{ourresult}) and in the two-point function of the circular Wilson loop with ${\cal O}_2$, eq. (\ref{cgrt}). The agreement between coefficients in eq. (\ref{powersym}) and in eq. (\ref{cgrt}) is perhaps not too surprising from the field theory point of view, since the radiated power can be read off from the stress-energy tensor, which as already mentioned is in the same short multiplet as ${\cal O}_2$, and the D3-brane corresponding to hyperbolic motion \cite{fiol11} is just the continuation to Lorentzian signature of the Euclidean D3-brane used to compute the vev of the circular Wilson loop \cite{Drukker:2005kx}. Let us nevertheless note that from the point of view of the D3-brane computations, this agreement was not a foregone conclusion, even if the D3-branes used in these three computations are related by a conformal transformation and/or continuation to Lorentzian signature. For instance, the energy loss computation in \cite{fiol11} captured physics of radiative fields, encoded in the bulk by the presence of a horizon in the world-volume  of the D3-brane, while in the computation of the previous section, the physics of static fields is captured by the behavior near the $AdS$ boundary, and the D3-brane world-volume has now no horizon.

Furthermore, this agreement between the coefficient of radiated power and the two point function of the circular Wilson loop with ${\cal O}_2$ also takes place for the antisymmetric representation, at least at the D-brane probe computation. This can easily be checked by comparing the relevant results obtained respectively in \cite{fiol11} and \cite{Giombi:2006de}, by means of D5-brane probes.

This second agreement in the coefficients of eq. (\ref{ourresult}) and eq. (\ref{powersym}), or perhaps more directly between eq. (\ref{powersym}) and eq. (\ref{cgrt}), leads us again to propose that given a particular representation of the gauge group, the full coefficients are exactly the same. In particular this proposal implies that the exact formula for radiated power of an infinitely heavy particle transforming in the fundamental representation and undergoing hyperbolic motion, valid for arbitrary values of N and $\lambda$ is
\begin{equation}
P=\frac{\lambda}{8\pi N}\frac{L_{N-1}^2(-\frac{\lambda}{4N})+L_{N-2}^2(-\frac{\lambda}{4N})}{L_{N-1}^1(-\frac{\lambda}{4N})}\frac{1}{R^2}
\label{resultwo}
\end{equation}
This is our second main result in this section. It reduces in the large $\lambda$, large $N$ limit to the one obtained by Mikhailov \cite{hep-th/0305196}. Furthermore, if we again set $k=1$ in the result of the D3-brane probe computation for the  k-symmetric representation, eq. (\ref{powersym}), we obtain an agreement in the large $\lambda$, large $N$, fixed $\kappa$ limit, even though $k=1$ is beyond the regime of validity of the computation \cite{fiol11} yielding (\ref{powersym}).

Let us conclude this paper by commenting on the possibility that the previous formula (\ref{resultwo}) for total radiated power might be valid not only for hyperbolic motion, but for an arbitrary timelike trajectory, with the obvious substitution $1/R^2 \rightarrow a^\mu a_\mu$. A first piece of evidence is that Mikhailov's computation \cite{hep-th/0305196} of the total radiated power for arbitrary timelike trajectories does indeed give a common $\sqrt{\lambda}$ coefficient, independent of the trajectory. Some further evidence beyond the leading large N, large $\sqrt{\lambda}$ result is given by the fact that for particles in the antisymmetric representation, the D5-brane computation in \cite{fiol11} again gives a Li\'enard-type formula for the total radiated power. This issue deserves further attention.

\section{Acknowledgements} 
We would like to thank Mariano Chernicoff for helpful conversations. We are particularly thankful to Nadav Drukker for bringing reference \cite{Giombi:2006de} to our attention, for urging us to compare our original results with the ones that appear in that reference, and for illuminating correspondence. The research of BF is supported by a Ram\'{o}n y Cajal fellowship, and also by MEC FPA2009-20807-C02-02, CPAN CSD2007-00042, within the Consolider-Ingenio2010 program, and AGAUR 2009SGR00168. The research of BG is supported by an ICC scholarship and by MEC FPA2009-20807-C02-02. The research of AL was partly supported by a Spanish MEPSYD fellowship for undergraduate students. AL further acknowledges support from Fundaci\'on Caja Madrid and the Perimeter Scholars International program.


\begin{thebibliography}{20}

\bibitem{Fiol:2011aa} 
  B.~Fiol, B.~Garolera and A.~Lewkowycz,
  ``Gluonic fields of a static particle to all orders in 1/N,''
  arXiv:1112.2345 [hep-th].

\bibitem{Dolan:2002zh} 
  F.~A.~Dolan and H.~Osborn,
  ``On short and semi-short representations for four-dimensional superconformal symmetry,''
  Annals Phys.\  {\bf 307}, 41 (2003)
  [hep-th/0209056].

\bibitem{Lee:1998bxa} 
  S.~Lee, S.~Minwalla, M.~Rangamani and N.~Seiberg,
  ``Three point functions of chiral operators in D = 4, N=4 SYM at large N,''
  Adv.\ Theor.\ Math.\ Phys.\  {\bf 2}, 697 (1998)
  [hep-th/9806074].

\bibitem{Baggio:2012rr} 
  M.~Baggio, J.~de Boer and K.~Papadodimas,
  ``A non-renormalization theorem for chiral primary 3-point functions,''
  arXiv:1203.1036 [hep-th].

\bibitem{Drukker:2000rr}
  J.~K.~Erickson, G.~W.~Semenoff and K.~Zarembo,
  ``Wilson loops in N=4 supersymmetric Yang-Mills theory,''
  Nucl.\ Phys.\ B {\bf 582}, 155 (2000)
  [hep-th/0003055].
  N.~Drukker, D.~J.~Gross,
  ``An Exact prediction of N=4 SUSYM theory for string theory,''
  J.\ Math.\ Phys.\  {\bf 42}, 2896-2914 (2001).
  [hep-th/0010274].
  V.~Pestun,
  ``Localization of gauge theory on a four-sphere and supersymmetric Wilson loops,''
  arXiv:0712.2824 [hep-th].

\bibitem{Drukker:2006ga} 
  N.~Drukker,
  ``1/4 BPS circular loops, unstable world-sheet instantons and the matrix model,''
  JHEP {\bf 0609}, 004 (2006)
  [hep-th/0605151].

\bibitem{Semenoff:2001xp} 
  G.~W.~Semenoff and K.~Zarembo,
 ``More exact predictions of SUSYM for string theory,''
  Nucl.\ Phys.\ B {\bf 616}, 34 (2001)
  [hep-th/0106015].
  
\bibitem{Okuyama:2006jc} 
  K.~Okuyama, G.~W.~Semenoff,
  ``Wilson loops in N=4 SYM and fermion droplets,''
  JHEP {\bf 0606}, 057 (2006).
  [hep-th/0604209].

\bibitem{Giombi:2006de} 
  S.~Giombi, R.~Ricci and D.~Trancanelli,
  ``Operator product expansion of higher rank Wilson loops from D-branes and matrix models,''
  JHEP {\bf 0610}, 045 (2006)
  [hep-th/0608077].

\bibitem{Aharony:1999ti} 
  O.~Aharony, S.~S.~Gubser, J.~M.~Maldacena, H.~Ooguri and Y.~Oz,
  ``Large N field theories, string theory and gravity,''
  Phys.\ Rept.\  {\bf 323}, 183 (2000)
  [hep-th/9905111].
E.~D'Hoker and D.~Z.~Freedman,
  ``Supersymmetric gauge theories and the AdS / CFT correspondence,''
  hep-th/0201253.

\bibitem{correa}
D.~Correa, J.~ Henn, J.~Maldacena, A.~Sever,
``An exact formula for the radiation of a moving quark in ${\cal N}=4$ super Yang-Mills,''
 arXiv:1202.4455 [hep-th].

\bibitem{fiol11} 
  B.~Fiol and B.~Garolera,
  ``Energy Loss of an Infinitely Massive Half-Bogomol'nyi-Prasad-Sommerfeld Particle by Radiation to All Orders in $1/N$,''
  Phys.\ Rev.\ Lett.\ \ {\bf 107}, 151601  (2011)
  [arXiv:1106.5418 [hep-th]].
  
\bibitem{Rey:1998ik}
  S.~-J.~Rey, J.~-T.~Yee,
  ``Macroscopic strings as heavy quarks in large N gauge theory and anti-de Sitter supergravity,''
  Eur.\ Phys.\ J.\  {\bf C22}, 379-394 (2001).
  [hep-th/9803001].

\bibitem{Maldacena:1998im}
  J.~M.~Maldacena,
  ``Wilson loops in large N field theories,''
  Phys.\ Rev.\ Lett.\  {\bf 80}, 4859-4862 (1998).
  [hep-th/9803002].


\bibitem{hep-th/9812007} 
  U.~H.~Danielsson, E.~Keski-Vakkuri and M.~Kruczenski,
  ``Vacua, propagators, and holographic probes in AdS / CFT,''
  JHEP\ {\bf 9901}, 002  (1999)
  [hep-th/9812007].

\bibitem{hep-th/9906153} 
  C.~G.~Callan, Jr. and A.~Guijosa,
  ``Undulating strings and gauge theory waves,''
  Nucl.\ Phys.\ B\ {\bf 565}, 157  (2000)
  [hep-th/9906153].

\bibitem{Athanasiou:2010pv}
  C.~Athanasiou, P.~M.~Chesler, H.~Liu, D.~Nickel, K.~Rajagopal,
  ``Synchrotron radiation in strongly coupled conformal field theories,''
  Phys.\ Rev.\  {\bf D81}, 126001 (2010).
  [arXiv:1001.3880 [hep-th]].
  Y.~Hatta, E.~Iancu, A.~H.~Mueller, D.~N.~Triantafyllopoulos,
  ``Aspects of the UV/IR correspondence : energy broadening and string fluctuations,''
  JHEP {\bf 1102}, 065 (2011).
  [arXiv:1011.3763 [hep-th]].
  Y.~Hatta, E.~Iancu, A.~H.~Mueller, D.~N.~Triantafyllopoulos,
  ``Radiation by a heavy quark in N=4 SYM at strong coupling,''
  Nucl.\ Phys.\ B {\bf 850}, 31 (2011)
  [arXiv:1102.0232 [hep-th]].
  R.~Baier,
  ``On radiation by a heavy quark in N = 4 SYM,''
  arXiv:1107.4250 [hep-th].

\bibitem{arXiv:1106.4059} 
  M.~Chernicoff, A.~Guijosa and J.~F.~Pedraza,
  ``The Gluonic Field of a Heavy Quark in Conformal Field Theories at Strong Coupling,''
  JHEP {\bf 1110}, 041 (2011)
  [arXiv:1106.4059 [hep-th]].
  
\bibitem{hep-th/0305196} 
  A.~Mikhailov,
  ``Nonlinear waves in AdS / CFT correspondence,''
  hep-th/0305196.

\bibitem{Forste:1999qn} 
  S.~Forste, D.~Ghoshal and S.~Theisen,
  ``Stringy corrections to the Wilson loop in N=4 superYang-Mills theory,''
  JHEP {\bf 9908}, 013 (1999)
  [hep-th/9903042].
 N.~Drukker, D.~J.~Gross and A.~A.~Tseytlin,
  ``Green-Schwarz string in AdS(5) x S**5: Semiclassical partition function,''
  JHEP {\bf 0004}, 021 (2000)
  [hep-th/0001204].
 V.~Forini,
  ``Quark-antiquark potential in AdS at one loop,''
  JHEP {\bf 1011}, 079 (2010)
  [arXiv:1009.3939 [hep-th]].
 N.~Drukker and V.~Forini,
  ``Generalized quark-antiquark potential at weak and strong coupling,''
  JHEP {\bf 1106}, 131 (2011)
  [arXiv:1105.5144 [hep-th]].

\bibitem{Gomis:2006sb}
  J.~Gomis, F.~Passerini,
  ``Holographic Wilson Loops,''
  JHEP {\bf 0608}, 074 (2006).
  [hep-th/0604007].
 ``Wilson Loops as D3-Branes,''
  JHEP\ {\bf 0701}, 097  (2007)
  [hep-th/0612022].

\bibitem{Yamaguchi:2006tq}
  S.~Yamaguchi,
  ``Wilson loops of anti-symmetric representation and D5-branes,''
  JHEP {\bf 0605}, 037 (2006).
  [hep-th/0603208].

\bibitem{Hartnoll:2006hr}
  S.~A.~Hartnoll, S.~Prem Kumar,
  ``Multiply wound Polyakov loops at strong coupling,''
  Phys.\ Rev.\  {\bf D74}, 026001 (2006).
  [hep-th/0603190].

\bibitem{Hartnoll:2006ib}
  S.~A.~Hartnoll,
  ``Two universal results for Wilson loops at strong coupling,''
  Phys.\ Rev.\  {\bf D74}, 066006 (2006).
  [hep-th/0606178].

\bibitem{hep-th/0104082} 
C.~G.~Callan, Jr., A.~Guijosa and K.~G.~Savvidy,
  ``Baryons and string creation from the five-brane world volume action,''
  Nucl.\ Phys.\ B\ {\bf 547}, 127  (1999)
  [hep-th/9810092].
 J.~M.~Camino, A.~Paredes and A.~V.~Ramallo,
  ``Stable wrapped branes,''
  JHEP\ {\bf 0105}, 011  (2001)
  [hep-th/0104082].

\bibitem{Hartnoll:2006is}
  S.~A.~Hartnoll, S.~Prem Kumar,
  ``Higher rank Wilson loops from a matrix model,''
  JHEP {\bf 0608}, 026 (2006).
  [hep-th/0605027].

\bibitem{Chernicoff:2006yp}
  M.~Chernicoff, A.~Guijosa,
  ``Energy Loss of Gluons, Baryons and k-Quarks in an N=4 SYM Plasma,''
  JHEP {\bf 0702}, 084 (2007).
  [hep-th/0611155].

\bibitem{Mueck:2010ja} 
  W.~Mueck,
  ``The Polyakov Loop of Anti-symmetric Representations as a Quantum Impurity Model,''
  Phys.\ Rev.\ D {\bf 83}, 066006 (2011)
  [Erratum-ibid.\ D {\bf 84}, 129903 (2011)]
  [arXiv:1012.1973 [hep-th]].
S.~Harrison, S.~Kachru and G.~Torroba,
  ``A maximally supersymmetric Kondo model,''
  arXiv:1110.5325 [hep-th].

\bibitem{Drukker:2005kx}
  N.~Drukker, B.~Fiol,
  ``All-genus calculation of Wilson loops using D-branes,''
  JHEP {\bf 0502}, 010 (2005).
  [hep-th/0501109].

\bibitem{hep-th/9702076} 
  I.~R.~Klebanov,
  ``World volume approach to absorption by nondilatonic branes,''
  Nucl.\ Phys.\ B\ {\bf 496}, 231  (1997)
  [hep-th/9702076].
  I.~R.~Klebanov, W.~Taylor and M.~Van Raamsdonk,
  ``Absorption of dilaton partial waves by D3-branes,''
  Nucl.\ Phys.\ B\ {\bf 560}, 207  (1999)
  [hep-th/9905174].

\bibitem{Berenstein:1998ij} 
  D.~E.~Berenstein, R.~Corrado, W.~Fischler and J.~M.~Maldacena,
  ``The Operator product expansion for Wilson loops and surfaces in the large N limit,''
  Phys.\ Rev.\ D {\bf 59}, 105023 (1999)
  [hep-th/9809188].
  
\end{thebibliography}
\end{document}